\documentclass[12pt]{article}

\usepackage{fancyhdr}
\usepackage{isomath}
\usepackage{amsmath}
\usepackage{amsbsy}
\usepackage{amssymb}
\usepackage{amscd}
\usepackage{amsfonts}
\usepackage{graphicx,color}
\usepackage{verbatim}
\usepackage{euscript}
\usepackage{stmaryrd}
\usepackage{subfigure}

\usepackage[margin=1in]{geometry}

\begin{document}
\title{A design principle for actuation of
nematic glass sheets}
\author{Amit Acharya\thanks{Dept. of Civil \& Environmental Engineering, and Center for Nonlinear Analysis, Carnegie Mellon University, Pittsburgh, PA 15213, email: acharyaamit@cmu.edu.}
}

\maketitle

\begin{abstract}
\noindent A continuum mechanical framework is developed for determining a) the class of stress-free deformed shapes and corresponding director distributions on the undeformed configuration of a nematic glass membrane that has a prescribed spontaneous stretch field and b) the class of undeformed configurations and corresponding director distributions on it resulting in  a stress-free \emph{given} deformed shape of a nematic glass sheet with a prescribed spontaneous stretch field. The proposed solution rests on an understanding of how the Lagrangian dyad of a deformation of a membrane maps into the Eulerian dyad in three dimensional ambient space. Interesting connections between these practical questions of design and the mathematical theory of isometric embeddings of manifolds, deformations between two prescribed Riemannian manifolds, and the slip-line theory of plasticity are pointed out.
\end{abstract}

\maketitle

\section{Introduction}
Design principles for obtaining desired shapes of thin bodies or slender bodies of soft materials in response to various physical stimuli has received much attention in the recent engineering, physics, and mathematics literature (e.g., \cite{Holmes, Nardinocci1, Nardinocci2}). Nematic glasses (explained in Sec. \ref{physics}) are a good specific example of such soft materials with potential for use in actuation. The basic engineering design goal of actuation of nematic glass sheets is to determine imprinted director distributions in a flat (or curved) sheet of nematic glass material, in order to achieve particular actuated shapes on exposure to appropriate stimuli.  Based on the creative fabrication techniques
of Broer, Sanchez-Somolinos and co-workers \cite{de2012engineering},
Modes, Warner and Bhattacharya \cite{modes2011blueprinting,
  modes2011gaussian}  \footnote{I would like to acknowledge Kaushik Bhattacharya
for introducing me to the subject of nematic glasses.}
have considered this question in some restricted special cases and in the context of determining the final deformed shapes when the patterned director field in the undeformed flat state is known. The inverse problem, that of determining the undeformed (unactuated) shape and a nematic director distribution on it that results in a known actuated shape, is significantly more difficult \cite{aharoni_sharon_kupferman,warner_mostajeran, plucinsky_lemm_bhatta} using the techniques that have been employed till date. In fact there does not exist a formulation whose solutions, even when assumed to exist, can be shown to solve the inverse problem in full generality.
  
In this note we propose a formulation, based essentially on an insight from continuum mechanics, that provides the governing equation for the whole class of actuated shapes and imprinting director distributions for a given, possibly spatially varying, opto-thermal stretch distribution. Moreover, the proposed formulation is in a sense `symmetric' for the forward and inverse design problems (see Sec. \ref{proposed_LCG} for definitions) so that it answers the design question for both problems when the prescribed opto-thermal stretch field is given.

In this connection, it is important to differentiate between the goals of the forward and inverse problems considered here and problems where a full metric field is specified (i.e., principal stretches as well as principal directions are specified, instead of just the principal stretches), along with either the undeformed or deformed configurations. When the full target metric is specified on the undeformed configuration, the problem of determining sufficiently smooth deformed configurations simply becomes that of the isometric embedding problem from 2 into 3 dimensions. Our forward problem has more freedom in that the principal directions are free to choose; it appears that this has not been noted and exploited in the literature thus far (cf. \cite{bhatta_lew_schaff} where $\mathrm{\Gamma}$-limits of elastic energy functionals for thin 3-d bodies containing prestrain fields relevant to our discussion are analyzed in the limit of vanishing thickness). This freedom has even more significance in the inverse problem. For the metric components specified on the manifold and the deformed configuration given, to our knowledge the only result available for solving the inverse problem exactly (and only  locally) is in  Theorem 8.1 in \cite{ach_lew_pak}. That result shows that the problem is strongly overconstrained, very different from the standard isometric embedding problem in its details, and solutions are expected to exist in rare circumstances (with a sufficient condition for existence of solutions  provided, not accessible to practical implementation; also note that the variational reformulation and analysis of the inverse problem with fully specified metric dealt with in \cite{ach_lew_pak} does not guarantee, by Theorem 2.4 and Corollary 3.4 and remarks immediately thereafter, that a solution to the inverse problem, i.e., an absolute minimum with value 0, exists for a given smooth metric field, either for the `3-d functional' or its `2-d' limit). However, the inverse problem for the prescribed principal stretch problem dealt with here is, as explained later, essentially the same as the forward problem. As noted already, this simplicity of the inverse problem has not been realized in the literature thus far. In addition, our analysis is not restricted to `plates', allowing for arbitrarily curved smooth reference membranes (defined in Sec. \ref{sec:prin_stretch}). 

\section{Physical basis of the actuation of nematic glass sheets} \label{physics}
We briefly review the constitution of soft solids with some amount of nematic order following \cite{avan2005large, Warner}, followed by the main physical idea behind their actuation. Nematic liquid crystals are made up of rod-like molecules; they have no positional ordering (like liquids) 
but have orientational ordering in that the rod-like molecules line up in some fixed orientation 
when not deformed by external stimuli. Any deviation from a uniformly oriented state costs energy 
and results in orientational elasticity.

Nematic liquid crystals elastomers (which we shall refer to by the shorthand LCE) are rubber-like, 
cross-linked, polymeric solids. Polymers are long-chain, flexible molecules, made of `monomer' constituents. 
Their extreme flexibility comes from the fact that, generally, a sequence of a more-or-less 
fixed number of monomers form rigid units, which are loosely jointed to form the long-chains. 
This flexibility allows polymers to flow under stress above their glass-transition temperature, 
i.e., in their fluid-like state. 
Elastomers (e.g., rubbers) are polymers that are cross-linked, i.e., the long chains are linked 
together at some points, thus giving the composite of many chains some rigidity. 
The shear modulus is much smaller ($\times 10^{-4}$) than in crystalline solids, 
while in bulk hydrostatic response they behave like incompressible liquids.

LCEs are elastomers whose polymer chains may consist of nematic liquid crystal monomers or they may 
be ordinary elastomers with liquid crystal molecules as `hanging' on pendant side-chains. 
Due to this detail of constitution, the average nematic orientation - the director - 
of a piece of LCE (even when not side-chain type) under homogeneous deformation, 
may not conform to the orientation of a material fiber fixed to the LCE matrix under the same 
applied deformation, when the material fiber and the director are aligned before deformation. 
Thus, LCEs have both positional elasticity - due to rubber-like, solid response of the polymer chains 
- and orientational elasticity due to the separately deforming director.

A nematic liquid crystal glass (LCG) is, in its turn,  a very highly cross-linked LCE such 
that the director is effectively constrained to move with the matrix. This is unless, possibly, 
when one is close to the defect cores, i.e., regions where there are severe changes in the director orientation. 
Essentially, nematic glasses are transversely isotropic solids and, for the problems to be dealt with in this note, 
they will be considered as elastic solids, often with immobile disclination defects in the nematic director 
field. Disclinations are locations where the director field is
discontinuous/contains localized high gradients. An LCG has a typical
elastic modulus (in the strong direction) of ~ 1 GPa (roughly a tenth
of a typical metal) while the LCE has a modulus of ~ 1 MPa. 

\subsection{Actuation of nematic glass
  sheets} \label{actuation} 
  
As explained in \cite{de2012engineering, modes2011blueprinting, modes2011gaussian}, the nematic mesogens, whether in LCE or LCG, are responsive to heat and light. 
When cooled, the mesogens in a chain tend to order and therefore `line-up' in a single orientation, 
thus having the effect of elongating the chain by uncoiling the spaghetti-like conformation.  
Heating has the opposite effect of increasing the disorder by a return to the coiled conformation, 
and thus shortening the chain. 
Similarly, exposure to light of particular wavelengths causes contraction or extension. 
When the mesoscale ordering is spatially homogeneous in a macroscopic body, heating or exposure 
to light cause macroscopic stretching/contractions of the body.

For the purposes of actuation of a sheet, one considers an ordered state to begin with, 
where the ordering may not be spatially homogeneous but patterned.
 Such patterning is induced by utilizing two glass substrates with thin  layers of liquid crystal polymer material, 
that have the intended patterns of the eventual sheet imprinted on them by the use of polarized ultra-violet light. 
The space in-between the layered substrates (cell) is then filled with a nematic liquid crystal material, 
that aligns itself to the surface by anchoring boundary conditions due to orientational elasticity. 
Then the whole cell is photopolymerized, i.e., the nematic liquid crystal is changed to a liquid crystal polymer 
through the action of light,
undergoing chain-forming reactions that keep the alignment of the liquid crystal state, as well 
as keeping the whole assembly heavily cross-linked. 
One now obtains the nematic glass sheet with a predetermined pattern on  extrication from the substrates.
Subsequently,  the sheet is exposed to heat or light. This causes further ordering of the polymerized chains 
along the directions of the imprinted local order, thus causing stretch along the pre-existing director direction 
and contraction in the direction orthogonal to it, and vice-versa for disordering. 
The mechanical effect on heating/light exposure can be summarized by a so-called `opto-thermal Poisson's ratio' 
$\nu$, characterizing the fact that if a stretch of magnitude $\lambda$ occurs in the ordering direction, 
then a stretch of magnitude $\lambda^{-\nu}$ occurs in all transverse directions to the director. 
This local deformation due to ordering/disordering from the patterned state is often referred to as 
\emph{spontaneous deformation}. If $n$ represents the director orientation in the sheet before 
being subjected to the stimuli, then the spontaneous deformation is given by the following tensor field 
on the unstimulated flat sheet:
\begin{equation}\label{nem} 
A_{nem} = \lambda^{-\nu}\, \mbox{Id}_2 + (\lambda - \lambda^{-\nu} )
n \otimes n,
\end{equation}
where $\mbox{Id}_2$ is the identity tensor in the plane of the reference sheet, $n$ depends upon position in the sheet, while $\lambda$ and $\nu$, in general, depend only 
on temperature or characteristics of incident light. 

In what follows, we will take the point of view that the unstimulated or undeformed membrane is not necessarily flat, since this generality is allowed by our mathematical formulation and leads to interesting applications \cite{modes2012responsive}. Let $\mathcal E_3$ be the ambient three-dimensional Euclidean point space, and the two-dimensional surface $U \subset \mathcal E_3$ the undeformed membrane. The elastic response functions of the membrane at any point $z \in U$ are assumed to be a function of $\tilde{F}(z) A_{nem}^{-1}(z)$, where $\tilde{F}(z)$ is the deformation gradient between the tangent space at $z$ of $U$ and the tangent space at $y(z)$ of the deformed membrane, where $y: U \rightarrow \mathcal E_3$ is the deformation of the membrane.  The director $n$ in \eqref{nem} corrresponding to $A_{nem}(z)$ for each $z \in U$ then belongs to the tangent space of $U$ at $z$ , and $\mbox{Id}_2$ there is the identity tensor on the same tangent space. Thus the energy density is assumed to be given by 
\begin{equation}\label{energy}
\psi( \tilde{F}(z) A_{nem}^{-1}(z) )
\end{equation}
(before restrictions due to frame-indifference are imposed).
The domain of the energy density function $\psi: \mathcal D \rightarrow \mathbb{R}$ is understood as follows: Consider the collection of the linear embeddings of the (2D) tangent space to $U$ at each of its points into the (3D) translation space of $\mathcal E_3$.
Consider now the set of orientation preserving invertible linear transformations between any two members of this collection. Then consider the set of all such invertible linear transformations generated from each such ordered pair of translation spaces. It is this last set of objects that is considered as the domain $\mathcal D$ of $\psi$. In addition, $\psi$ is assumed to have the property that it attains a minimum value if and only if its argument is of the form $R$, where $R$ is a rotation (proper orthogonal) tensor between any two of the translation spaces involved in the definition of $\mathcal D$.  

For the purpose of finding equilibrium configurations of the sheet, $A_{nem}(z)$ will be considered as a given field (unlike in the nematic elastomer case). Note that $A_{nem}$ is analogous to the plastic
distortion tensor $F^p$ in elastoplasticity theory (see, e.g., \cite{turzi_et_al, turzi} for applications to soft materials), but now in the context of 2-d bodies occupying non-Euclidean subsets of $\mathcal  E_3$; the main modeling hypothesis is that, to the extent allowed by compatibility and
displacement boundary conditions, the `total' deformation gradient $\tilde{F}$ field attempts to generate a right stretch tensor field that equals that of $A_{nem}$ pointwise in order to attain minimum energy in the body.

It is important to note that in each local region of the sheet, this ordering/disordering of the chain 
microstructure causes no stress in the region, if not constrained by the ordering/disordering in adjoining regions. 
This observation forms the physical basis of \emph{actuation}. The predetermined pattern in the sheet combined 
with the stimuli-induced stretching may set up a local spontaneous stretch/metric field with nontrivial 
Gaussian curvature. 
If the material of the sheet has to be stress-free, then it needs to take up this spontaneous stretch of the LCG 
and therefore cannot remain in-plane. Bending out-of-plane may allow such stress-free deformations, 
with a much lower bending penalty, due to thinness of the sheet. 
Clearly, if no bending out-of plane can accommodate the spontaneous deformation field, then further distortion 
of local material elements takes place beyond the spontaneous deformation, 
resulting in higher energy states of the sheet. 

The formulation proposed in this note capitalizes on the physical details of opto-thermal stimulation of nematic glass materials and the geometry of sheets to identify non-trivial sheet deformations that have no membrane stretching energy regardless of the frame-indifferent constitutive response used for membrane deformation. A small amount of bending energy may be involved and this we account for as a higher order effect in our design problems, but exactly.

\section{Design problems in actuation of
  nematic glass sheets}\label{proposed_LCG} 

The intuition behind the posed design problems is as follows. 
Let $\sigma_i$, $i=1,2$ be two prescribed scalar fields.
Suppose we are successful in finding a \emph{compatible} symmetric, positive definite tensor
field $C$ on the undeformed membrane $U\subset \mathcal E_3$, whose tensor square root has the form
$$ \sqrt{C(x)} = \sigma_1(x) d(x) \otimes d(x) + \sigma_2(x) d^{\perp} (x) \otimes d^{\perp} (x),$$
where $d(x), d^\perp(x) \in T_xU$ are two unit vectors belonging to the tangent space of $U$ at each $x \in U$, which are mutually perpendicular.
By `compatible', we mean: 
\begin{itemize}
\item in the language of mechanics, that there exists a deformation of the undeformed membrane into 3d space whose Right Cauchy Green tensor field is
given by $C$ (with right stretch tensor field $\sqrt{C}$). 
\item in the language of differential geometry, the components of $C$ viewed as those of a Riemannian metric specified as a function of coordinates parametrizing $U$ has an isometric embedding into $\mathbb{R}^3$. 
\end{itemize} 
Physically, the $\sigma_i$-s could be the moduli $\{ \lambda,
\lambda^{-\nu} \}$ in (\ref{nem}), defined 
as spontaneous principal stretches in response to the opto-thermal stimuli.

If we now `pattern in,' or `write' into the tangent spaces of $U$ the field $n(x) = d(x)$ for each 
$x\in U$ \footnote{It is easiest to think of this `writing' when $U$ is flat.} using the field $d$ determined above, then $C$ can be viewed as the right Cauchy Green tensor
field of a stress-free deformation $\phi: U \rightarrow \mathcal E_3$ of the undeformed membrane with deformation gradient $\tilde{F}$ after opto-thermal stimulation, i.e.,
$\tilde{F}^T\tilde{F} = C$. Further, $d$ and $d^{\perp}$ 
would be then the principal directions of the stretch tensor of this deformation, 
stretched by the prescribed amounts $\sigma_i$ and remaining mutually
perpendicular, i.e., $(\tilde{F}d) \cdot (\tilde{F}d^\perp) = 0$,
 even though generally having moved out of the reference plane: $(\tilde{F}d)\cdot M \neq 0$, where $M$ is the unit normal field on the reference membrane.

Thus, the whole problem of inducing stress free deformations, or in other words \emph{actuation}, 
due to opto-thermal stimuli of the undeformed membrane reduces to:
\begin{itemize}
\item[(i)] finding compatible stretch tensor fields with prescribed principal stretches,
\item[(ii)] for any one of these fields, doing a pointwise spectral decomposition 
to find the principal direction fields,
\item[(iii)] writing in the nematic director field in the undeformed configuration along one of the
principal direction fields.
\end{itemize}

It is perhaps important here to mention the advantage of nematic glass material over a nematic elastomer 
for such actuation. Considering a flat undeformed membrane, suppose the opto-thermal stretch field is such that no in-plane compatible deformation can accommodate it as  a principal stretch field, but out-of-plane deformations can. In the LCG, energy minimization would then force an out-of-plane deformation. However, in the LCE, it is possible that the energy can be minimized by director \emph{reorientation} accompanying in-plane material deformation of the rubber matrix.

In the following,
\begin{itemize}
\item  we refer to the question of determining
  deformed (actuated) shapes of the membrane, given the undeformed (unactuated) configuration
  and the prescribed spontaneous stretches, as the {\bf forward design problem}. 
\item The {\bf inverse design problem} refers to the question of
  determining the undeformed (unactuated) shape, given the deformed (actuated) configuration of the
  membrane and the spontaneous prescribed stretches. 
\end{itemize}
In both cases, the imprinting director distribution on the undeformed configuration is determined as in (ii) and (iii) above.

\subsection{A membrane problem with prescribed principal stretches}\label{sec:prin_stretch}

In solving the forward problem, a surface $S$ (the deformed membrane to be solved for) constitutes the \emph{target} with
a given surface $U$ (the undeformed membrane) as the \emph{reference}. For the inverse
problem, $U$ (the undeformed membrane) will be the target and a given $S$ (the deformed membrane) the reference.  \emph{In the following, we will refer to the target membrane as $T$ and the reference membrane as $R$.} We will also assume that $(\tilde{\lambda}_1, \tilde{\lambda}_2)$ is a given pair of positive real number fields on the reference $R$ with $\tilde{\lambda}_1 \geq \tilde{\lambda}_2$; they could be formed from the sets of generalized spontaneous principal stretches $\left\{
  \sigma_1, \sigma_2 \right\}$ or $\left\{ \frac{1}{\sigma_2},
  \frac{1}{\sigma_1} \right\}$ (depending on whether we want to solve
the forward or the inverse problem).

Denote a generic point on the reference membrane as $X = \sum_{i = 1}^3 X_i c_i$, where $(c_i, i = 1,2,3)$ is an orthonormal basis for $\mathcal{E}_3$, and a generic point on
the target as $x= \sum_{i = 1}^3 x_i c_i$. In the following, we will \emph{invoke the summation convention, unless otherwise mentioned.} Latin indices will be range over $(1,2,3)$ and Greek indices over $(1,2)$. 

When there exists a deformation that maps the reference onto the target membrane, we refer to its deformation gradient as $F$ and the 
Right Cauchy Green tensor of the deformation map as $C = F^T F$. For the forward problem, we obtain the desired imprinting pattern
directly from the eigenframe field of $C$. For the inverse problem, we
map the eigenframe field of $C$ by $F$ (recall the definitions of the target and reference membranes for the inverse problem) and normalize to obtain
the pattern on $U$. 

With respect to a local convected coordinate parametrization 
$\xi = (\xi^1, \xi^2)$ of the reference and the target (the existence of which implicitly induces a deformation between the two), the natural bases for the coordinate system on the reference and
target membranes are given by $ E_\alpha := \partial_\alpha X$ and
$\partial_\alpha x$, respectively. Defining $G_{\alpha \beta} = \partial_\alpha X
\cdot \partial_\beta X$ and $G^{\alpha \beta} = \left[G_{\alpha
    \beta} \right]^{-1}$ as its matrix inverse, with the $\cdot$ representing the dot product of two vectors in $\mathcal{E}_3$, the dual basis on the
reference membrane is $E^\alpha =  G^{\alpha
  \beta} E_{\beta}$. Then $F = \partial_\alpha x \otimes
E^\alpha$ and
\begin{equation}\label{eqn:RCG}
C = F^T F = g_{\alpha \beta} E^\alpha \otimes E^\beta
\end{equation}
with $g_{\alpha \beta} := \partial_\alpha x
\cdot \partial_\beta x.$ 
The characteristic equation of the tensor $C$ is 
\begin{equation}\label{eqn:charac}
\mu^2 - tr(C) \mu + det(C) = 0,
\end{equation}
where
$$tr(C) = C : \mbox{Id} =
g_{\alpha \beta} E^\alpha \otimes E^\beta :
E_\rho \otimes E^\rho =  g_{\alpha
  \beta}\delta^\beta_\rho G^{\alpha \rho} =
g_{\alpha \rho}G^{\alpha \rho}$$ 
and
$$ det(C) = \left(
  \frac{| \partial_1 x \times \partial_2 x|}{| \partial_1 X
    \times \partial_2 X|} \right)^2 = \frac{det(g)}{det(G)},$$
where $g,G$ are the matrices defined earlier. On solving
for the two ordered roots of \eqref{eqn:charac}, the governing equation for the deformation map $x$ is given by the following nonlinear, first-order, system
\begin{equation}\label{eqn:embed}
\begin{split}
tr(C) + \sqrt{ (tr(C))^2 - 4 det(C)} = 2 \left( \tilde{\lambda_1} \right)^2 \\
tr(C) - \sqrt{ (tr(C))^2 - 4 det(C)} = 2 \left(\tilde{\lambda}_2\right)^2
\end{split}
\end{equation}
of 2 scalar partial differential equations for 3 fields defining the components of $x(\xi) = x^i(\xi) c_i$.

In the case of nematic glass sheets, the prescribed stretches are unequal and constant on $U$. The formulation is general for the principal stretches prescribed as  two fields $\tilde{\lambda}_i (X(\xi)), i = 1,2$. At any point where the prescribed stretches were to be equal, the pair of governing equations would change to
\[
tr(C) = 2 {\tilde{\lambda}}^2
\]
\[
det(C) = {\tilde{\lambda}}^4,
\]
where $\tilde{\lambda}$ is the common value of the prescribed principal stretch. It is clear that it is best to consider problems where the specified principal stretch fields are such that they are everywhere equal or everywhere unequal.
\subsubsection{Global $C^1$ and local $C^3$ solutions}\label{sec:C_1_3}
We assume the reference membrane $R$ to be a compact, orientable surface; for simplicity we assume that $R$ can be parametrized by a single coordinate patch $D \subset \mathbb{R}^2$ and the parametrization $X:D \rightarrow \mathcal{E}_3$ is of class $C^1$ (so that any such parametrization is as well, by the standard smoothness between coordinate patches). Consider now \emph{any} pair of orthonormal, tangent vector fields, say $P = (P_1, P_2)$, that is H\"{o}lder continuous on $R$. We assume that the prescribed principal stretch fields are  H\"{o}lder continuous on $R$ as well. Then define
\begin{equation}\label{eqn:def_metric}
g_{\alpha \beta} (X; P) = E_{\alpha}(X) \cdot \left [ \sum_{i=1}^{2} \left( \tilde{\lambda}_i(X) \right)^2 P_i(X) \otimes P_i (X) \right]  E_{\beta} (X), \ \ X \in R,
\end{equation}
and in the following we simply refer to the metric field so defined as $g_{\alpha \beta} (P)$. Due to the fact that $R$ is a smooth embedded surface in $\mathcal{E}_3$ (with metric field $G_{\alpha \beta}$ that is H\"{o}lder continuous), Theorem 2 of the review \cite{avner_friedman} describing the results of Nash \cite{nash_c1} (improved by Kuiper) states  that there exists an isometric embedding with metric $g_{\alpha \beta} (P)$, which in our setting may be interpreted as the existence of a $C^1$ map $x: D \rightarrow \mathcal{E}_3$ satisfying $\partial_\alpha x \cdot \partial_\beta x = g_{\alpha \beta} (P)$.
In fact, due to the compactness of $R$ in $\mathcal{E}_3$ and the continuity of $g_{\alpha \beta}(P)$ on it, a \emph{short}  $C^1$ embedding w.r.t. the metric $g_{\alpha \beta}(P)$ can be produced by an appropriate constant scaling of the function $X: D \rightarrow \mathcal{E}_3$, i.e., if $\bar{X}$ is the short embedding, then $\partial_\alpha \bar{X} \partial_\beta \bar{X} \leq g_{\alpha \beta}$ in the sense of quadratic forms. Assuming that the H\"{o}lder exponent of the $g_{\alpha \beta}(P)$ field is $\geq \frac{2}{19}$, Corollary 2 and Remark 1 of Conti, De Lellis, and Szekelyhidi \cite{CO_DL_SZ} implies that there exists an \emph{infinite} number of $C^{1,\alpha}, \alpha < \frac{1}{19}$, isometric embeddings  $x: D \rightarrow \mathcal{E}_3$, each satisfying $\partial_\alpha x \cdot \partial_\beta x = g_{\alpha \beta} (P)$, for each fixed choice of the pair $P$. The vast collection of isometric embeddings arises from the (proven) property \cite{CO_DL_SZ} that the short embedding can be uniformly approximated by the $C^{1,\alpha}$ embeddings on $D$, i.e., $|| x - \bar{X}||_{C^0} < \epsilon$ for any $\epsilon > 0$.

Thus it is clear that the forward \emph{and} inverse design problems actually have a huge class of global solutions (when the prospect of varying the pair $P$ is also taken into account).

In contrast to the abundant existence of $C^1$ global isometric embeddings corresponding to a given metric field, the known existence of embeddings of 2-d surfaces in 3-d space with higher regularity, particularly of class $C^3$, for metrics with Gauss curvature fields with indefinite sign is scarce. Theorem 1.1 of Han \cite{han} is typical of such \emph{local} results, a part of which states that if $g$ is a $C^9$ metric field whose Gauss curvature vanishes at some point of a coordinate neighborhood and its gradient does not vanish at the point, then $g$ admits a $C^3$ isometric embedding in a neighborhood of that point. It is also classically known that two $C^2$ compact surfaces with identical metric fields with positive Gauss curvature can at most differ by a rigid deformation \cite{CO_DL_SZ}.

Thus, again, an infinite number of local solutions of higher regularity (class $C^3$) of the forward and inverse design problems are guaranteed by utilizing the result just mentioned \cite{han} by using the metric defined in \eqref{eqn:def_metric}, but now with the $P$ and $\tilde{\lambda}$ fields chosen to be of appropriately high regularity.

\subsection{A prescribed principal stretch problem from one given surface into another}\label{sec:defmn_riem}
The considerations of Sec. \ref{sec:C_1_3} show that in fact there is a great deal of freedom in defining deformations of a given surface with prescribed principal stretches. Given this information, which is not necessarily convenient for practical design considerations due to the massive non-uniqueness, it is natural to explore whether prescribed principal stretch deformations can be defined between two arbitrarily chosen smooth surfaces. It turns out that this problem has been studied in great detail, in the local context, by Deturck and Yang \cite{det-yang} and Gevirtz and Chuaqui \cite{Ge1, Ge2} (\emph{these mathematical results require the strict inequality  $\tilde{\lambda}_1 > \tilde{\lambda}_2$}). They show, for varying hypotheses, that the problem of local existence of deformations with prescribed principal stretches between two equidimensional Riemannian manifolds leads to a study of diagonal hyperbolic systems for which unique local solutions exist for appropriately specified Cauchy data. In the following, we make the minor contribution of developing the explicit form of the governing $2 \times 2$ nonlinear hyperbolic system that represents the deformation of an open set of a given surface into another specified surface, satisfying the constraint that it has prescribed principal stretch fields, meant for the practitioner of continuum mechanics not familiar with the language of modern differential geometry.

Let the mappings $X:D \rightarrow \mathcal{E}_3$ and $x: d \rightarrow \mathcal{E}_3$ be parametric representations of two \emph{given} surfaces $R, T$, respectively, where $D,d \subset \mathbb{R}^2$ are open sets. Thus, $R = \{ X(\xi): \xi \in D \}$ and $T = \{ x(\eta): \eta \in d \}$. We define generic points in $D$ as $\xi = (\xi^1, \xi^2)$ and $\eta = (\eta^1, \eta^2)$, respectively. With some abuse of notation, we look for a differentiable injective map $\eta: D \rightarrow d$ with pointwise non-vanishing determinant, the existence of which implicitly induces a deformation $X \mapsto x(X)$ of $R$ through the composition $x \circ \eta \circ X^{-1}$ ( where $X^{-1}$ is now interpreted as a function on $X(D)$). We use the definition $\frac{\partial x}{\partial \eta^\alpha} = e_\alpha : d \rightarrow V_3$ ($V_3$ is the translation space of $\mathcal{E}_3$), with $h_{\alpha \beta} (\eta) = e_\alpha(\eta) \cdot e_\beta(\eta)$.

We define the two-point tensor
\[
F(\xi) = \frac{\partial \eta^\alpha}{\partial \xi^\beta} (\xi) \, e_\alpha (\eta(\xi)) \otimes E^\beta (\xi)
\]
(the deformation `gradient') from the tangent space at $X(\xi) \in R$ to the tangent space at $x(\eta(\xi)) \in T$ and the Right Cauchy-Green tensor
\begin{equation}\label{eqn:FtF}
\begin{split}
& F^T F (\xi) = \frac{\partial \eta^\alpha}{\partial \xi^\beta} (\xi) \, h_{\alpha \gamma} (\eta(\xi)) \, \frac{\partial \eta^\gamma}{\partial \xi^\delta} (\xi)\, E^\beta (\xi) \otimes E^{\delta} (\xi) =: g_{\beta \delta}(\xi)\, E^\beta (\xi) \otimes E^{\delta} (\xi);\\
& g_{\beta \delta} \left( \frac{\partial \eta}{\partial \xi}, \eta \right)= \frac{\partial \eta^\alpha}{\partial \xi^\beta} \, h_{\alpha \gamma} (\eta) \, \frac{\partial \eta^\gamma}{\partial \xi^\delta}
\end{split}
\end{equation}
from the tangent space at $X(\xi) \in R$ to itself. Substituting this expression for $g$ (in terms of $\eta$ and the derivatives of the mapping $\eta$) into \eqref{eqn:RCG}, \eqref{eqn:charac}, and \eqref{eqn:embed} we obtain the explicit expression for the $2 \times 2$ nonlinear, first-order system for the function $\eta: D \rightarrow d$ (cf. \cite{warner_mostajeran}). As already mentioned,  the theory of local solutions to this system is well understood through the works \cite{det-yang, Ge1, Ge2}. Very interestingly, there is a direct connection between this formulation of the problem and problems of slip-line fields in the theory of plasticity \cite {Ge1,Ge3}.

\subsection{A variational basis for computational approximation}\label{sec:numerics}
In order to generate computational approximations of the target membrane given the the reference, we propose the following functional of $x:R \rightarrow \mathcal{E}_3$:
\begin{equation}\label{eqn:main_eqn}
L(x) = \int_R
\left[ \left| \mu_1(F)  - (\tilde{\lambda}_1)^2 \right|^p + \left|
    \mu_2(F) - (\tilde{\lambda}_2)^2 \right|^p \right] ~\mbox{d}a,
\end{equation}
where $p > 0$ is a constant with $p = 2$ a natural choice, $R$ is the reference membrane and $da$ is the area measure on it. Recall that $F (Y) = \partial_\alpha (x \circ X )(X^{-1}(Y)) \otimes E^\alpha(Y), Y\in R$, for $X(\cdot)$ being any parametrization of $R$. Any absolute (global) minimizer of $L$ (that have been shown already to exist) yields the solution of the prescribed principal stretch mapping we seek. The solution of the problem automatically yields, by a `post-processing' step, the required imprinting director distribution, as explained in Sec. \ref{sec:prin_stretch}. The minimization problem remains valid for the determination of approximately stress-free deformed membranes with corresponding imprinting director distributions in the undeformed membrane.

The considerations in Sec. \ref{proposed_LCG} indicate that  solutions to the minimization problem, i.e., ${\sf{argmin}} \, L(x)$, suffer from `massive' nonuniqueness.  This observation, coupled with the considerations of Sec. \ref{sec:defmn_riem} suggests that the practical way to seek approximate solutions to the problem is to consider an ($L^2$) gradient flow of the `energy' \eqref{eqn:main_eqn}:
\[
\partial_t \,x = -  \frac{\delta L}{\delta x} (x),
\]
where $t$ is a fictitious time-scale and the right-hand-side of the equation is the variational derivative of $L$, and look for its equilibria starting from appropriate initial conditions. These initial conditions can be set to be any desired smooth target shape to be achieved, and the observations in Sec \ref{sec:defmn_riem} suggest that, at least locally, one can expect to achieve such targets. The variational formulation allows significant liberty with respect to regularity and global shapes may as well be attempted to be approached by the same procedure. Of course, minimal boundary conditions need to be imposed to
eliminate rigid modes, i.e., solutions that differ from each other by
rigid body deformations. 
  
The determination of such deformations (and imprinting director field) may be, for example, approached via a finite-element method based approximation scheme by discretizing the reference membrane by finite elements in the usual way, beginning with a `membrane-only' implementation. Let $R^h = \cup_i \, A_i$ be the discrete rendition of the reference membrane as the union of finite elements, each $A_i$ 
parametrized by local coordinates coordinates
$\xi^\alpha, \alpha = 1,2$ forming the set
$[-1,1] \times [-1,1] =: \Box$. Let $N^A(\xi^1,\xi^2),
A = 1 \ldots T$ be finite element test and trial functions,
where $T$ corresponds to the number of nodes in the mesh. Then we
approximate positions on the target and reference membranes as 
\begin{equation}\label{eqn:isoparam}
x^h = \sum_{A,i} x^A_i N^A c_i, \quad X^h = \sum_{A,i} X^A_i N^A c_i,
\end{equation}
where $x^A_i$ and $X^A_i$, $A = 1\ldots T, i = 1\ldots 3$ are the
global Cartesian coordinates of the position of the $A^{th}$ node of
the mesh on the corresponding membranes. It is a property of the isoparametric finite element representation \eqref{eqn:isoparam} that the function $x^h$, appropriately interpreted, ends up being globally $C^0$ and piecewise $C^{\infty}$ on $R^h$. The corresponding approximate functional (on a finite-dimensional space) becomes

\[
L^h(x^h) = \int_{R^h}
\left[ \left( \mu_1(F^h) - \left( \tilde{\lambda}_1 \right)^2 \right)^p + \left( \mu_2(F^h) - \left( \tilde{\lambda}_2 \right)^2 \right)^p \right] \ \mbox{d}a^h
\]
where $F^h = \partial_\alpha x^h \otimes \left(E^h\right)^\alpha$ (no sum on $h$, of course) and $\left(E^h\right)^\alpha$ is the dual basis of $\partial_\alpha X^h$, and $\mbox{d}a^h$ is the area measure on $R^h$. The contribution to $L^h$ from each such element is given by
\begin{equation}\label{num_func}
\begin{split}
\int_{\Box} \left[ \left( \mu_1(F^h(X^h(\xi)) - \left( \tilde{\lambda}_1 (X^h(\xi)) \right)^2 \right)^p  + \left( \mu_2(F^h(X^h(\xi))  - \left( \tilde{\lambda}_2(X^h(\xi)) \right)^2 \right)^p \right] \\
| \partial_1 X^h (\xi)
\times \partial_2 X^h(\xi)| ~\mbox{d}\xi^1 \mbox{d}\xi^2.
\end{split}
\end{equation} 

Our general strategy accommodates curved undeformed
membranes with non-trivial topology (e.g., non-simply connected undeformed membranes), thus allowing the study, in  generality, of the interaction of defects in patterning with these aspects of the undeformed membrane as explained in \cite{modes2012responsive} . 

\subsection{A bending regularization}\label{sec:bending}
As already mentioned in Sections \ref{actuation} and \ref{proposed_LCG}, the proposed formulation specific to the optothermal stimulation of nematic glass sheets is guaranteed to result in vanishing membrane stretching energy by design, independent of the membrane constitutive response, beyond frame-indifference. Of course, the spontaneous principal stretches $\tilde{\lambda}_1, \tilde{\lambda}_2$ are specific to the constitution of the nematic glass material and thereby will affect the class of actuated shapes produced for each specific material. What we have outlined up until now is certainly the dominant part of the stated design problems. As a refinement, and to mitigate the lack of uniqueness of solutions (as may be expected from the available results for  $C^2$ embeddings with positive Gauss curvature and the general $C^3$ isometric embeddings discussed in Sec. \ref{sec:C_1_3}), it is natural to consider bending deformations, which is a higher order, small energy effect compared to membrane stretching due to thinness of the sheets that, nevertheless, can induce greater regularity in solutions. In essence, if the mathematical model does not contain any bending energy penalty, the membrane stress-free shapes can occur with sharp ridges and singular points where bending deformation, described by an appropriate function of the second fundamental form of the deformed shape, can be singular. In physical reality, such extreme bending coupled with the the thinness of the sheet will produce a small amount of bending energy around such singular lines and points. To account for this refinement, it is natural to include a bending penalty in the governing functional \eqref{eqn:main_eqn} in terms of a bending stiffness times an appropriate function of the deformation gradient $F$, and the second-fundamental forms $\mbox{grad}\, m := \partial_\alpha m \otimes e^\alpha$ and $\mbox{Grad}\, M := \partial_\alpha M \otimes E^\alpha$ where $m$ and $M$ are the unit normal field on the target and reference, respectively, and $e^\alpha = F^{-T} E^\alpha$. In choosing a natural candidate for this bending penalty function, one may consider the fact that the change in the second fundamental form of a surface penalizes deformations that may not be physically related to bending (e.g., radial expansion of a right-circular cylinder), and therefore use physically appropriate and kinematically exact bending strain measures \cite{acharya2000nonlinear}. Of course, with such a bending regularization numerical approximations would require either a mixed formulation with the normal field interpolated as a separate field or techniques like isogeometric analysis. 
\section{Concluding remarks}
The main contribution of this note is to develop a \emph{common} mathematical framework for solving the forward and inverse problems of design of thin sheets subjected to opto-thermal stretching. That such a `symmetric' framework can exist may be considered surprising based on  works in the existing literature, e.g., \cite[Sec. 6.2]{plucinsky_lemm_bhatta} and \cite[Sec. 1]{warner_mostajeran}. A critical realization is that approaching the design problem of opto-thermal stretching of membranes by considering the full metric as specified is sub-optimal, leading to particularly severe, and unnecessary, difficulties in answering the question of determining the undeformed geometry and director distribution on it, given the deformed membrane.

With respect to practicality of the design principle developed in the paper, there are two types of questions. One relates to computing approximate solutions to the forward and inverse design  problems; some of these issues have been dealt with in Sections \ref{sec:numerics} and \ref{sec:bending}, with backing from rigorous mathematical results. Since the interest here is definitely in global minimizers, and it has already been established that (without bending regularization) uniqueness of solutions is absent in a large class of functions that are less than $C^2$ regular, using the technique of $\mathrm{\Gamma}$-convergence on approximate finite-element solution sequences parametrized by the mesh size appears worth pursuing. In addition, there is a vast literature in plasticity theory and optimal design where questions of computing slip-line fields and Hencky-Prandtl nets arise; it is natural to explore to what extent these approaches can be adapted to the problem posed in Sec. \ref{sec:defmn_riem}. 

The other question relates to how uniquely can the desired shapes be realized on actual, physical, actuation with the computationally determined director distributions imprinted in the undeformed configuration. This can only be answered with certainty on practical testing, but the `rigidity' in obtained shapes facilitated by smooth profiles due to \emph{physical} bending energy cost - that is expected to be invariably present in small amounts in real, thin membranes - can only be expected to help.

\section*{Acknowledgment}
It is a pleasure to acknowledge discussions with Julian Gevirtz, Reza Pakzad, and Marta Lewicka. I would like to thank an anonymous referee for various comments that have improved the presentation of the paper.


\begin{thebibliography}{9999}

\bibitem{Holmes} M. Pezzulla, S. A. Shillig, P. Nardinocchi, and D. P. Holmes, \textit{Morphing of geometric composites via residual swelling}, Soft Matter {\bf 11} no. 29 (2015), 5812--5820.

\bibitem{Nardinocci1} P. Nardinocchi, L. Teresi, and V. Varano, \textit{Strain induced shape formation in fibred cylindrical tubes}, Journal of the Mechanics and Physics of Solids {\bf 60} no. 8 (2012), 1420--1431.

\bibitem{Nardinocci2} L. Lucantonio, P. Nardinocchi, and M. Pezzulla, \textit{Swelling-induced and controlled curving in layered gel beams}, Proceedings of the Royal Society of London A: Mathematical, Physical and Engineering Sciences, {\bf 470} no. 2171 (2018) 20140467.

\bibitem{avan2005large} K. D. Harris, R. Cuypers, P. Scheible,
  C. L. van Oosten, C. W. M. Bastiaansen, J. Lub and D. J. Broer,
  \textit{Large amplitude light-induced motion in high elastic modulus
    polymer actuators}, Journal of Materials Chemistry, {\bf 15}
  (2005), 5043--5048. 
  
\bibitem{de2012engineering} L. T. de Haan, C. Sanchez-Somolinos,
  C. M. W. Bastiaansen, A. P. H. J. Schenning and D. J. Broer,
  \textit{Engineering of complex order and the macroscopic deformation
    of liquid crystal polymer networks}, Angewandte Chemie
  International Edition, {\bf 51} (2012), 12469--12472.  

\bibitem{modes2011blueprinting} C.D. Modes and M. Warner,
  \textit{Blueprinting nematic glass: Systematically constructing and
    combining active points of curvature for emergent morphology},
  {\bf 84} (2011), 021711-1--7. 
  
\bibitem{modes2011gaussian} C.D. Modes, K. Bhattacharya and
  M. Warner, \textit{Gaussian curvature from flat elastica sheets},
  Proceedings of the Royal Society A: Mathematical, Physical and
  Engineering Science, {\bf 467} (2011), 1121--1140. 
  
\bibitem{aharoni_sharon_kupferman} H. Aharoni, E. Sharon, R. Kupferman, \textit{Geometry of Thin Nematic Elastomer Sheets}, Phys. Rev. Lett., {\bf 113}, no. 25, (2014), 257801.

\bibitem{warner_mostajeran} M. Warner, C. Mostajeran, \textit{Mapping director fields to metric variation, Gaussian curvature and topography}, arXiv preprint arXiv:1712.03136, (2017).

\bibitem{plucinsky_lemm_bhatta} P. Plucinsky, M. Lemm, and K. Bhattacharya, \textit{Actuation of thin nematic elastomer sheets with controlled heterogeneity}, Archive for Rational Mechanics and Analysis {\bf 227} no. 1 (2018), 149--214.

\bibitem{bhatta_lew_schaff} K. Bhattacharya, M. Lewicka, and M. Sch\"{a}ffner, \textit{Plates with incompatible prestrain}, Archive for Rational Mechanics and Analysis {\bf 221} no. 1 (2016), 143--181.

\bibitem{ach_lew_pak} A. Acharya, M. Lewicka, and M. R. Pakzad, \textit{The metric-restricted inverse design problem}, Nonlinearity {\bf 29} no. 6 (2016), 1769-1797.

\bibitem{turzi_et_al} P. Biscari, A. DiCarlo, and S. S. Turzi, \textit{Liquid relaxation: A new Parodi-like relation for nematic liquid crystals}, Physical Review E {\bf 93} no. 5 (2016), 052704.

\bibitem{turzi} S. S. Turzi, \textit{Active nematic gels as active relaxing solids}, Physical Review E {\bf 96} no. 5 (2017), 052603.

\bibitem{Warner} M. Warner and E. Terentjev, \textit{Liquid crystal elastomers},
  Oxford University Press (2003).
  
\bibitem{avner_friedman} A. Friedman, \textit{Isometric embedding of Riemannian manifolds into Euclidean spaces}, Reviews of Modern Physics {\bf 37} no. 1 (1965), 201--203.

\bibitem{nash_c1} J. F. Nash, \textit{$C^1$ Isometric imbeddings}, Annals of Mathematics, {\bf 60} no. 3 (1954), 383-396.

\bibitem{CO_DL_SZ} S. Conti, C. De Lellis, and L. Székelyhidi Jr., \textit{h-principle and rigidity for $C^{1,\alpha}$ isometric embeddings}, Nonlinear Partial Differential Equations, Abel Symposia 7, (2012), ed. H. Holden and K. H. Karlsen, 83--116.

\bibitem{han} Q. Han, \textit{On the isometric embedding of surfaces with Gauss curvature changing sign cleanly}, Communications on Pure and Applied Mathematics {\bf 58} no. 2 (2005), 285--295.

\bibitem{det-yang} D. M. DeTurck, and D. Yang, \textit{Existence of elastic deformations with prescribed principal strains and triply orthogonal systems}, Duke Mathematical Journal {\bf 51} no. 2 (1984), 243--260.

\bibitem{Ge1} M. Chuaqui, and J. Gevirtz, \textit{Constant principal strain mappings on 2-manifolds}, SIAM Journal on Mathematical Analysis {\bf 32} no. 4 (2000), 734--759.

\bibitem{Ge2} J. Gevirtz, \textit{A diagonal hyperbolic system for mappings with prescribed principal strains}, Journal of Mathematical Analysis and Applications {\bf 176} no. 2 (1993), 390--403.

\bibitem{Ge3} J. Gevirtz, \textit{On planar mappings with prescribed principal strains}, Archive for Rational Mechanics and Analysis {\bf 117} no. 4 (1992), 295--320.

\bibitem{modes2012responsive} C.D. Modes and M. Warner,
  \textit{Responsive nematic solid shells: Topology, compatibility,
    and shape}, Europhysics Letters, {\bf 97} (2012), 36007-1--4. 

\bibitem{acharya2000nonlinear} A. Acharya, \textit{A nonlinear
    generalization of the Koiter--Sanders--Budiansky bending strain
    measure}, International Journal of Solids and Structures, {\bf 37}
  (2000), 5517--5528.
  


  
\end{thebibliography}
\end{document}